\documentclass[%
 reprint,
 amsmath,amssymb,
 aps,
]{revtex4-2}
\usepackage{amssymb}

\usepackage{graphicx}
\usepackage{dcolumn}
\usepackage{bm}
\usepackage{color}

\newcommand{\dd}{{\rm d}}

\newcommand{\ee}{{\rm e}}
\newcommand{\beq}{\begin{equation}}
\newcommand{\eeq}{\end{equation}}

\newcommand{\Hy}{H$_2$ }

\newcommand{\ltsim}{\lesssim}

\newcommand{\Schm}{\text{\textit{Sc}}}

\newcommand{\Pe}{\text{\textit{Pe}}}
\newcommand{\Reyn}{\text{\textit{Re}}}
\newcommand{\Sh}{\text{\textit{Sh}}}
\newcommand{\p}{\partial}
\newcommand{\bnabla}{\bm{\nabla}}

\newcommand{\review}[1]{{\color{black}{#1}}}

\begin{document}

\preprint{APS/123-QED}

\title{Advection-modulated gaseous diffusion through an orifice}

\author{Mario Sánchez Sanz}
 \email{mssanz@ing.uc3m.es}
 \affiliation{Department of Thermal and Fluids Engineering, Carlos III University, Madrid, 28930 (Spain)}
\author{Antonio~L.~S\'anchez }%
\affiliation{Department of Mechanical and Aerospace Engineering, UC San Diego, La Jolla,
                   92161 (USA)} 

\date{\today}

\begin{abstract}

We examine flow and transport through an orifice in a flat wall separating semi-infinite atmospheres of two dissimilar gases. The analysis assumes steady conditions and order-unity values of the Schmidt number $\Schm$ and Péclet number $\Pe$, such that advection and diffusion contribute comparably to mass and momentum transport. Mixing between the two gases induces order-unity variations in viscosity and density, resulting in coupled concentration and velocity fields. The solution yields the mass transfer rates of both gases, expressed in terms of an appropriately defined Sherwood number, as well as the overpressure required to sustain the flow, all as functions of $\Schm$ and $\Pe$. An explicit analytical solution is obtained in the limit of small $\Pe$, while numerical integration is used to describe flows with $\Pe = O(1)$. The mixing of hydrogen and air is used as an illustrative example that serves to highlight the influence of large gas-molecular-weight differences on the flow structure and associated mixing rate, with additional selected results given for the case of hydrogen and water vapor.

\end{abstract}

\maketitle


\section{Introduction}

As a canonical configuration representative of restrictive flow orifices, also known as flow restrictors, we consider the flow of a fluid through a circular orifice in a thin wall separating semi-infinite regions filled with two different fluids. The mass flux across the orifice $\dot{m}$ depends on its radius, the pressure difference between the two atmospheres, and the physical properties of the fluids on either side.
Orifices provide a compact and reliable means of flow control without the need for moving parts, maintain a stable flow rate and reduce fluctuations that can occur due to pressure variations in the supply lines \review{\cite{reader2015orifice}.}

Most studies in the literature focus on liquid transfer through an orifice as a means to regulate mass flow rate \cite{johansen1930flow}. This configuration has wide-ranging applications, including refrigeration systems \cite{tu2006refrigerant}, cavitation studies \cite{mishra2005cavitation}, liquid mixing \cite{luo2013experimental}, and microparticle focusing \cite{park2009continuous}. The first theoretical analysis of this problem was conducted by Sampson \cite{sampson1891}, who considered low-Reynolds-number flow through a circular orifice in an infinitesimally thin wall. By formulating the problem in oblate-spheroidal coordinates, he derived an analytical solution for the velocity field and the associated pressure loss. This work was later extended by Dagan et al. \cite{dagan1982infinite}, who developed creeping-flow solutions for orifices in walls of small but finite thickness. Sampson’s analysis was further generalized to hourglass-shaped orifices using numerical simulations by Gravelle et al. \cite{Gravelle2013,Gravelle2014}. On the experimental side, Johansen \cite{johansen1930flow} was the first to measure the pressure drop through a small confined orifice, providing a way to quantify the discharge coefficient as a function of Reynolds number.

The mixing of two different liquids was recently studied by Atwal et al. \cite{atwal2022mass}, who leveraged the fact that liquid diffusivities are typically several orders of magnitude smaller than their kinematic viscosities. Consequently, momentum transport is far more effective than mass transport, resulting in flows where viscous forces dominate--i.e., flows characterized by a small Reynolds number--but advection and diffusion contribute comparably to scalar transport--i.e., species transport characterized by an order-unity Péclet number. In this regime, the hydrodynamics decouple from the transport problem, allowing the use of previously derived analytical velocity fields \cite{sampson1891,dagan1982infinite} to evaluate the advective transport rate. This leads to a linear transport problem, which Atwal et al. \cite{atwal2022mass} solve approximately to yield the mass transfer rate as a function of the Péclet number. A related configuration, incorporating fluid–wall interactions, was examined by Rankin et al. \cite{rankin2019entrance}.

\begin{figure}[!ht]
    \centering
    \includegraphics[width=\linewidth]{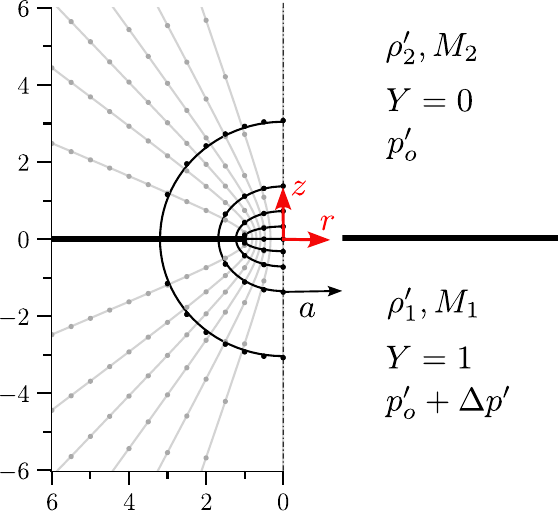}
    \caption{A schematic representation of the orifice-flow configuration studied, including an indication of the cylindrical coordinates used in the description. The left half of the plot show streamlines $\psi=$constant (gray solid curves, $\Delta \psi = 0.025$ ) and isocontours of mass fraction $Y$ (black solid curves, $\Delta Y=0.1$) calculated numerically for $Pe=\alpha=\beta=0$. Dots are used to represent the analytical solutions given in Eqs.~\eqref{psi_sampson} and~\eqref{Y_sampson}.}
    \label{fig:sketch}
\end{figure}

Flow restrictors to regulate the mass flow of a gas stored in a gas reservoir into another gas has an inherent interest in several industries and scientific applications, such as flow visualization \cite{park2009effect}, instrument calibration \cite{oliveira2010numerical}, safety equipment \cite{shrouf2003pressure} and semiconductors fabrication \cite{ohkawa2002highly}. To examine this scenario in greater detail, we consider the specific problem of a gas, with density $\rho'_1$, \review{viscosity $\mu'_1$, and pressure $p'_0+\Delta p'$, flowing through an orifice of radius $a$ into a semi-infinite domain filled with a different gas with density $\rho'_2$, viscosity $\mu'_2$ and pressure $p'_0$, as sketched in Fig.~\ref{fig:sketch}. It is assumed that the overpressure $\Delta p'$ driving the motion satisfies $\Delta p' \ll p'_0$, resulting a low-Mach-number flow with steady mass flux $\dot{m}$}. Since the density and viscosity of the gas mixture across the fluid domain exhibits changes of order unity, the velocity and composition fields are strongly coupled, complicating the analysis. An additional difference from the liquid-flow case addressed in \cite{atwal2022mass} is that, for gases, the binary diffusion coefficient $D$ is typically comparable to the kinematic viscosities, $\mu'_1/\rho'_1 \sim \mu'_2/\rho'_2$, leading to order-unity values of the Schmidt number
\beq \label{Sc_def}
\Schm=\frac{\mu'_1/\rho'_1}{D}\sim \frac{\mu'_2/\rho'_2}{D}.
\eeq
For example, the Schmidt number associated with the binary diffusion coefficient $D$ of hydrogen in air is approximately $\mathrm{Sc} \approx 0.24$ or $\mathrm{Sc} \approx 1.66$ when using the properties of air or hydrogen, respectively, to evaluate the kinematic viscosity $\mu'_1/\rho'_1$.
Since the molecular transport rates of momentum and species are comparable, the Péclet number,
\beq \label{Pe_def}
\Pe=\frac{\dot{m}}{\rho'_1 a D},
\eeq
defined here using the density $\rho'_1$ of gas 1, is of the same order as the Reynolds number $\Reyn = \Pe/\Schm = \dot{m}/(\mu'_1 a)$. Thus, the simplifications used in the liquid case \cite{atwal2022mass}, which rely on the separation $\Reyn \ll \Pe \sim 1$, do not apply here. Instead, \review{to investigate configurations in which advection and diffusion jointly control the mass transfer rate,} the analysis must address the more general regime $\Reyn \sim \Pe \sim 1$, with flow velocities, of order $(\mu'_1/\rho'_1)/a$, that are comparable to the diffusion velocities, of order $D/a$.

In typical applications, the orifice radius $a$ is small but remains larger than the mean free path $\lambda$, ensuring a small Knudsen number $\lambda/a \ll 1$ and the validity of the continuum approximation used below. Cases where $a \ltsim \lambda$ fall outside this regime and require a kinetic-theory treatment, as in the work of Willis \cite{willis1965mass} for large Knudsen numbers $\lambda/a \gg 1$, and more recently in studies by Bykov and Zakharov \cite{bykov2020binary,bykov2022rarefied} on binary-gas outflow into vacuum through orifices. Departures from the continuum description will not be considered in the following analysis, which also neglects buoyancy effects, an approximation justified when the orifice radius satisfies $a \ll (\mu'_1/\rho'_1)^{2/3}/g^{1/3}$. Under this condition, the buoyancy-induced velocity \review{in the orifice region (i.e., at distances from the orifice comparable to its radius)}, of order $g a^{2}/(\mu'_1/\rho'_1)$, is much smaller than the pressure-driven velocity $(\mu'_1/\rho'_1)/a$ characteristic of the regime $\Pe \sim \Reyn \sim 1$ considered here. \review{For larger Reynolds numbers $\Reyn \gg 1$, the residence time in the orifice region decreases, and the criterion for neglecting buoyancy becomes less restrictive; specifically, buoyancy may be neglected provided $a \ll \Reyn^{2/3} (\mu'_1/\rho'_1)^{2/3}/g^{1/3}$, as follows from comparing the characteristic buoyancy-induced velocity $g a^{2}/[\Reyn (\mu'_1/\rho'_1)]$ with the pressure-driven velocity $\Reyn (\mu'_1/\rho'_1)/a$. Although buoyancy effects may eventually become relevant sufficiently far downstream in such high-Reynolds-number jets, the associated modifications of the far-field velocity have a negligible influence on the rate of gas transport through the orifice, thereby justifying their neglect in the present analysis.}

The rest of the paper is organized as follows. The dimensionless formulation needed to determine the mass transfer rate of each gas through the orifice as a function of the P\'eclet number is presented in section 2, followed in section 3 by the analytic solution corresponding to $\Pe \ll 1$. The general solution corresponding to $\Pe \sim 1$ is obtained numerically in section 4, with results given for selected gas pairs \review{over a range of P\'eclet numbers, from the purely diffusive limit $\Pe=0$ to moderately large values ($\Pe=12$) for which advection dominates mass transfer through the orifice.} The solution includes the computation of the small pressure difference needed to sustain the mass flow rate. Finally, concluding remarks are given in section 5.

\section{Formulation} \label{sec: formulation}

\subsection{Problem description}

Consider two dissimilar gases at the same temperature $T_o$ separated by an infinitesimally thin impermeable wall containing a communicating orifice of radius $a$ through which the two gases may inter-diffuse. The composition of the binary mixture is to be described in terms of the mass fraction $Y$ of gas $1$, with the mass fraction of gas 2 being simply $1-Y$. As indicated in the schematic shown in Fig.~\ref{fig:sketch}, the resulting steady solution is to be described using cylindrical coordinates $(r',z')$ with origin at the orifice center and corresponding velocity $\bm{v}'=(v'_r,v'_z)$. For definiteness, we shall assume that the ambient pressure  $p'_o+\Delta p'$ in the semi-space $z<0$, occupied by gas 1, is slightly larger than the ambient value $p'_o$ found on the other side, so that gas flows across the orifice in the positive $z$ direction.

We focus on configurations with $\Delta p' \ll p'_o$, such that the induced velocities remain much smaller than the speed of sound, resulting in a low-Mach-number flow that modulates the intermixing of the two gases. The associated flow rate $\dot{m} >0$ can be computed by evaluating the mass flux across the orifice
\beq \label{m_eq}
\dot{m} = \int_0^a \left( \rho' v'_z \right)_{z'=0} 2 \pi r' {\rm d} r',
\eeq
where $\rho'$ is the mixture density, which can be computed in terms of $Y$ with use of the low-Mach-number form of the equation of state 
\beq \label{eos}
\rho'=\frac{p'_o/(R^0 T_ o)}{Y/M_1+(1-Y)/M_2}.
\eeq
Here, $R^0$ is the universal gas constant and $M_1$ and $M_2$ are the molecular masses of the two gases, whose densities satisfy $\rho'_2/\rho'_1=M_2/M_1$. Density changes across the orifice can be significant when the two gases have very different molecular mass. For instance, in problems involving hydrogen and air and hydrogen and water vapor the density ratio is $\rho'_{\rm AIR}/\rho'_{\rm H_2}=M_{\rm AIR}/M_{\rm H_2} \simeq 15$  and $\rho'_{\rm H_2O}/\rho'_{\rm H_2}= M_{\rm H_2O}/M_{\rm H_2} \simeq 9$, respectively, with even higher values achieved when the heavy gas is either sulfur hexaflouride ($M_{\rm SF_6}=146.06$ g/mol) or uranium hexaflouride ($M_{\rm UF_6}=352.02$ g/mol).

With \review{barodiffusion} having a negligible effect in the limit $\Delta p' \ll p'_o$ considered here, the multicomponent transport diffusion equation \cite{hirschfelder1964molecular} can be seen to reduce {\it exactly} \cite{williams1985combustion} to the familiar Fick's law $\bm{V}_1 Y=-\bm{V}_2 (1-Y)=-D \bnabla' Y$, where $\bm{V}_1$ and $\bm{V}_2$ are the diffusion velocities of the two gases and $D$ is their binary diffusion coefficient, with $\bnabla'=(\p /\p r',\p /\p z')$ denoting the gradient operator. Our analysis assumes that the P\'eclet number, defined above in Eq.~\eqref{Pe_def}, takes order-unity values, so that the overall transport rate of each gas component through the orifice, $\dot{m}_1$ and $\dot{m}_2=\dot{m}-\dot{m}_1$, is a result of the competition of advection and diffusion, as described by
\beq \label{m1m2_eq}
\dot{m}_1=\dot{m}-\dot{m}_2=\int_0^a \left(\rho' Y v'_z - \rho' D \frac{\p Y}{\p z'}\right)_{z'=0} 2 \pi r' {\rm d} r'.
\eeq

The transport problem delineated above can be posed as that of finding the mass flux $\dot{m}$ and the transport rates $\dot{m}_1=\dot{m}-\dot{m}_2$ corresponding to a given pressure difference $\Delta p'$. An alternative procedure that simplifies the solution consists of selecting the mass flux $\dot{m}$, and therefore the value of the P\'eclet number, defined in Eq.~\eqref{Pe_def}, and determining the corresponding transport rates $\dot{m}_1=\dot{m}-\dot{m}_2$ and overpressure $\Delta p'$, giving the results in an appropriately nondimensional form. This latter approach is to be followed below.

\subsection{Dimensionless formulation}

The problem is formulated in dimensionless form using $a$ and $u_c=\dot{m}/(\rho'_1 a^2)$ as characteristic scales of length and velocity to define dimensionless variables $\bm{r}=(r,z)=(r'/a,z'/a)$ and $\bm{v}=(v_r,v_z)=(v'_r/u_c,v'_z/u_c)$, so that Eq.~\eqref{m_eq} reduces to
\beq \label{m_eq2}
\int_0^1 \left(\rho v_z\right)_{z=0} 2 \pi r {\rm d} r=1,
\eeq
where $\rho=\rho'/\rho'_1$. In terms of these variables, the conservation equations take the reduced form
\begin{align}
\bnabla \cdot (\rho \bm{v})&=0, \label{cont} \\
(\Pe/\Schm) \, \rho \, \bm{v} \cdot \bnabla \bm{v}&=-\bnabla p + \bnabla \cdot \bm{\tau}, \label{mom}\\
\Pe \, \rho \, \bm{v} \cdot \bnabla Y  &= \bnabla \cdot (\rho \bnabla Y), \label{Yeq}
\end{align}
where $\bnabla=(\p /\p r,\p /\p z)$. Here, $\tau=\mu (\nabla \bm{v}+\nabla^{{\rm T}} \bm{v})$ is the non-isotropic component of the viscous stress tensor scaled with $\mu'_1 u_c/a$, with $\mu=\mu'/\mu'_1$ representing the dimensionless viscosity coefficient of the gas mixture. Correspondingly, 
\beq \label{p_def}
p=\frac{p'-p'_o +(2/3) \mu' \bnabla' \cdot \bm{v}'}{\mu'_1 u_c/a} 
\eeq
is the sum of the spatial pressure difference $p'-p'_o$ and the isotropic component of the viscous stress tensor $(2/3) \mu' \bnabla' \cdot \bm{v}'$, both scaled with \textcolor{black}{$\mu'_1 u_c/a$}. The above equations must be supplemented with the equation of state~\eqref{eos} written in the dimensionless form
\beq \label{eos2}
\rho=\frac{1}{1+\alpha (1-Y)}, \quad {\rm where} \quad \alpha=\frac{M_1}{M_2}-1,
\eeq
along with an equation describing the variation of the dimensionless viscosity $\mu(Y)=\mu'/\mu'_1$ with the mixture composition. In the following, we adopt for simplicity the so-called Graham's model \citep{graham1846xxviii}, which assumes that the viscosity of the gas mixture is the sum of the products of the viscosities of their individual components and their mole fractions, resulting in the compact equation
\beq \label{mu_eq}
\mu=\rho [1+\beta (1-Y)], \quad {\rm where} \quad 
\beta=\frac{\mu'_2}{\mu'_1} \frac{M_1}{M_2}-1,
\eeq
which naturally reduces to $\mu=1$ when $\mu'_2=\mu'_1$ and $M_1=M_2$. More accurate expressions for $\mu(Y)$ could be easily incorporated in the formulation at the expense of increasing the number of parameters. 

Unlike the constant-property case $\alpha = \beta = 0$ shown in figure \ref{fig:sketch}, where perfect symmetry around the plane of the orifice is preserved in the case $Pe=0$, the dependency of both density and viscosity with concentration will break the symmetry about $z=0$ even when inertial effects are absent. 

Equations~\eqref{cont}--\eqref{Yeq} must be integrated subject to the boundary conditions
\beq \label{bc1}
v_r=v_z=\frac{\p Y}{\p z}=0 \quad {\rm at} \quad z=0 \quad {\rm for} \quad r>1,
\eeq
as corresponds to a no-slip impermeable wall, along with the symmetry conditions
\beq \label{bc2}
v_r=\frac{\p v_z}{\p r}=\frac{\p Y}{\p r}=0 \quad {\rm at} \quad r=0 \quad {\rm for} \quad -\infty < z < \infty,
\eeq
and the ambient conditions
\begin{align} 
Y \rightarrow 0, \quad &\frac{v_r}{r}=\frac{v_z}{z} \rightarrow \frac{3}{2 \pi} \frac{z^2}{(r^2+z^2)^{5/2}}  \quad {\rm as} \quad  r^2+z^2 \rightarrow \infty \label{bc3} 
\end{align}
for $z>0$ and
\begin{align}
Y \rightarrow 1, \quad &\frac{v_r}{r}=\frac{v_z}{z} \rightarrow -\frac{3}{2 \pi} \frac{z^2}{(r^2+z^2)^{5/2}} \quad {\rm as} \quad  r^2+z^2 \rightarrow \infty , \label{bc4}
\end{align}
for $z<0$, 
where it is assumed that the velocity approaches far from the orifice the Stokes-flow solution developed by Harrison~\cite{harrison1920pressure} (see also \cite{gusarov2020entrainment}), consistent with a velocity decaying with the inverse square of the distance.

\subsection{Preliminary considerations}

The integration of Eqs.~\eqref{cont}-\eqref{Yeq} supplemented by Eqs.~\eqref{eos2} and~\eqref{mu_eq} with the boundary conditions shown in Eqs.~\eqref{bc1}--\eqref{bc4} determines the velocity and composition fields everywhere. The solution across the orifice can be used to evaluate the transport rates of both gases, which can be conveniently expressed in terms of the Sherwood number
\beq \label{Sh_def}
\Sh=\frac{\dot{m}_1}{\rho'_1 a D}=\int_0^1 \left(\Pe \, \rho \, Y v_z - \rho \frac{\p Y}{\p z}\right)_{z=0} 2 \pi r {\rm d} r,
\eeq
as follows from Eq.~\eqref{m1m2_eq}.

As previously mentioned in the introduction, in liquid mixtures, addressed in~\cite{atwal2022mass,rankin2019entrance}, the Schmidt number takes extremely large values often exceeding one thousand, with the result that the Reynolds number $\Reyn=\Pe/\Schm$ is much smaller than the P\'eclet number. Under those conditions, the velocity field is described by the creeping-flow solution through an orifice, which can be expressed in close analytical form~\citep{sampson1891}. This simplification leads, upon substitution into Eq.~\eqref{Yeq}, to a linear problem for the mass fraction that can be approximately solved using series expansion in terms of Legendre polynomials~\citep{atwal2022mass}. By way of contrast, for the gas mixtures considered here, the Schmidt number is typically of order unity. In this regime, advection plays a central role in the transport of both momentum and species when $\Pe \sim 1$, resulting in a fundamentally nonlinear problem that is further complicated by the order-unity variations of density and viscosity with the mass fraction.

Although numerical integration is in general needed to evaluate the solution, analytical descriptions are available in some cases of interest. The simplest case arises when the P\'eclet number is very large $\Pe \rightarrow  \infty$. In this limit, the species conservation equation~\eqref{Yeq} reduces to $\bm{v} \cdot \bnabla Y$, indicating that $Y$ is conserved along streamlines, giving $Y=1$ at the orifice. The Sherwood number then reduces to $Sh=Pe$ as follows from Eq.~\eqref{Sh_def} when Eq.~\eqref{m_eq2} is taken into account, corresponding to the trivial case $\dot{m}_1=\dot{m}$ and $\dot{m}_2=0$. The opposite case $\Pe \ll 1$ of weak advection is also amenable to an analytical description, as described below.

\section{The limit $\Pe \ll 1$}

At leading order, in the limit $\Pe \ll 1$, the composition is determined by integration of $\bnabla \cdot (\rho \bnabla Y)=0$ supplemented by Eq.~\eqref{eos2} with the boundary conditions stated in Eqs.~\eqref{bc1}--\eqref{bc3}. In the constant-density case $\rho=1$, the problem can be reduced to that describing the potential surrounding an electrified disk, solved analitically by Weber~\cite{weber}. The resulting mass fraction can be expressed in the form 
\beq
Y=\frac{1}{\pi} \int_0^\infty \zeta^{-1} \sin \zeta \, {\rm e}^{-\zeta z} J_0(\zeta r) {\rm d} \zeta,
\eeq 
where $J_0$ is the Bessel function of the first kind of order zero and $\zeta$ is a dummy integration variable, yielding
\beq \label{Sh_constant density}
\Sh=-\int_0^1 \left. \frac{\p Y}{\p z}\right|_{z=0} 2 \pi r {\rm d} r=2 \quad {\rm for} \quad \Pe \ll 1.
\eeq
for the Sherwood number.

In solving the variable density case, it is convenient to introduce oblate spheroidal coordinates $(\xi,\eta)$~\cite{sneddon1966mixed}, defined such that
\beq
r=\cosh(\xi) \cos(\eta) \quad {\rm and} \quad z=\sinh(\xi) \sin(\eta).
\eeq 
In terms of these coordinates, also used in the previous analysis of liquid mixing~\citep{atwal2022mass}, the problem reduces to that of integrating 
\beq
\frac{1}{\cosh(\xi)} \frac{\p}{\p \xi}\left[\cosh(\xi) \rho(Y) \frac{\p Y}{\p \xi}\right]+\frac{1}{\cos(\eta)} \frac{\p}{\p \eta}\left[\cos(\eta) \rho(Y) \frac{\p Y}{\p \eta}\right]=0
\eeq
for $-\infty < \xi < \infty$ and $0\le \eta \le \pi/2$ with boundary conditions
\begin{align}
& \frac{\p Y}{\p \eta}=0 \; {\rm at} \; \eta=0,\frac{\pi}{2} \\ 
& Y\rightarrow 0 \; {\rm as} \; \xi \rightarrow \infty, \ {\rm and} \ \ Y\rightarrow 1 \; {\rm as} \; \xi \rightarrow -\infty,
\end{align}
as follows from Eqs.~\eqref{bc1}--\eqref{bc3}. 

The above boundary conditions are consistent with a solution independent of $\eta$, leading to the boundary-value problem
\beq \label{BV_problem}
\frac{\rm d}{{\rm d} \xi}\left[\cosh(\xi) \rho(Y) \frac{{\rm d} Y}{{\rm d} \xi}\right]=0 \quad \left\{\begin{array}{lll} Y\rightarrow 0 & {\rm as} & \xi \rightarrow \infty \\ Y\rightarrow 1 & {\rm as} & \xi \rightarrow -\infty \end{array} \right.,
\eeq
which can be integrated to give
\beq \label{Y_Pe<<1}
Y=1-\frac{1}{\alpha} \left\{\exp\left[\frac{2}{\pi}\ln(1+\alpha) \tan^{-1}(\ee^\xi)\right]-1 \right\}.
\eeq
The above expression~\eqref{Y_Pe<<1} can be written in cylindrical coordinates, yielding 
\beq \label{paraboloidal surfaces}
\sin^2\left[\pi \frac{\ln[1+\alpha(1-Y)]}{\ln(1+\alpha)}\right] r^2+ \tan^2\left[\pi \frac{\ln[1+\alpha(1-Y)]}{\ln(1+\alpha)}\right] z^2=1.
\eeq
for the isocontours of mass fraction. In representing the paraboloidal surfaces~\eqref{paraboloidal surfaces}, one must select $z<0$ for $Y>Y_o$ and $z>0$ for $Y<Y_o$, with
\begin{equation}
Y_o=1-\frac{1}{\alpha} \left(\sqrt{1+\alpha}-1 \right), \label{eq:Y_o}
\end{equation}
denoting the uniform value of the mass fraction across the orifice, obtained by evaluating Eq.~\eqref{Y_Pe<<1} at $\xi=0$. The correspoding density $\rho(0)=1/(\sqrt{1+\alpha})$, obtained from~Eq.~\eqref{eos2}, can be expressed in dimensional form to give
\beq
\rho'=\sqrt{\rho'_1 \rho'_2} \quad {\rm at} \quad z'=0 \quad {\rm for} \quad 0 < r' \le a, 
\eeq
indicating that in this limit $\Pe \ll 1$ the density at the orifice is the geometrical mean of the densities of the two gases.

Across the orifice surface $z=0$ with $0 \le r \le 1$, corresponding to $\xi=0$ and $\pi/2 \ge \eta \ge 0$, the diffusive flux can be evaluated using $\rho \p Y/\p z=\ln(1+\alpha)/(\alpha \pi \sin \eta)$ and $r=\cos \eta$ to give
\begin{align}
& -\int_0^1  \rho \frac{\p Y}{\p z} 2 \pi r {\rm d} r = \\-&\int_0^{\pi/2} 2 \pi \frac{1}{\alpha \pi} \ln(1+\alpha) \cos \eta \, \dd \eta = 
 \frac{2}{\alpha} \ln(1+\alpha), \nonumber
\end{align}
which naturally reduces to the constant-density result given in Eq.~\eqref{Sh_constant density} when $\alpha \ll 1$. Since the mass fraction across the orifice, given in Eq.~\eqref{eq:Y_o}, is uniform, it is straightforward to use Eq.~\eqref{m_eq2} to obtain the accompanying advective flux
\beq
\Pe \int_0^1 \rho Y v_z 2 \pi r {\rm d} r=\Pe \, Y_o,
\eeq
finally giving
\beq \label{eq:Sh}
\Sh=\frac{2}{\alpha} \ln(1+\alpha)+\left[1-\frac{1}{\alpha} \left(\sqrt{1+\alpha}-1 \right)\right] \Pe
\eeq
as an expression for the Sherwood number for $\Pe \ll 1$. Notice that $\Sh =2+Pe/2$ as $\alpha \to 0$.

This approximate solution, independent of $\Schm$ and $\beta$, corresponds to the first two terms in a rigorous asymptotic development of $Sh$ for $\Pe \ll 1$, involving expansion of the different variables in powers of $\Pe$. The computation of corrections of order $\Pe^2$ and higher, carrying dependences on $\Schm$ and $\beta$, requires quantification of the velocity field by integration of Eq.~\eqref{mom}, with the first term in the solution for the velocity corresponding to creeping flow, to be computed with account taken of the variations of density and viscosity associated with the nonuniform mass fraction shown in Eq.~\eqref{Y_Pe<<1}. Since these corrections are relatively small, this higher-order analysis is not further pursued here.

\section{Selected numerical results}

For $\Pe \sim 1$, the solution must be obtained numerically. Inertia breaks the symmetry present at low Péclet numbers, rendering oblate spheroidal coordinates less advantageous. Consequently, the simulations are performed using the governing equations in cylindrical coordinates $(r, z)$.

For a given Péclet number, determined by the specified mass flux $\dot{m}$, the solution depends on the gas pair through the values of $S_c$, $\alpha$, and $\beta$. Table~\ref{table} lists these parameters for H$_2$–Air and H$_2$–H$_2$O mixtures, of inherent industrial interest \cite{taniguchi1981growth, irene1982role}. Each pair yields two distinct sets of values, depending on which gas is at the higher pressure (designated as gas 1 in our notation). 

Note that in studying the mixing of air and hydrogen, air can be treated as a single gaseous species rather than as a mixture, since its two main components, N$_2$ and O$_2$, have similar molecular properties. In particular, their binary diffusion coefficients with H$_2$ differ by only about 10\%. Because their rates of diffusion into hydrogen are nearly identical, the relative N$_2$-to-O$_2$ ratio remains essentially uniform throughout the flow. As a result, the composition of the mixture can be characterized by a single mixture fraction $Y$, using mixture-averaged properties for air: a molecular mass $M_{\rm Air} \simeq 28.97$ g/mol and a binary diffusivity with hydrogen $D_{\rm H_2-Air} \simeq 0.674$ cm$^2$/s at standard temperature and pressure.


\begin{table}
\begin{center}
\begin{tabular}{cccccc}
       \hline
     & H$_2$-Air & Air-H$_2$ & H$_2$-H$_2$O  & H$_2$O-H$_2$ \\
    \hline
    $\beta$   & -0.85 & 5.73 & -0.87 & 6.80 \\
    $\alpha$  & -0.93 & 13.00 & -0.89 & 8.00 \\
    $Sc$      &  1.60& 0.25	& 1.43	& 0.20\\
    \hline
\end{tabular}
\label{table}
\caption{Set of parameters for the representative cases studied in this work.} 
\end{center}
\end{table}

\subsection{Numerical method}

\begin{figure}[!ht]
    \centering
    \includegraphics[width=\linewidth]{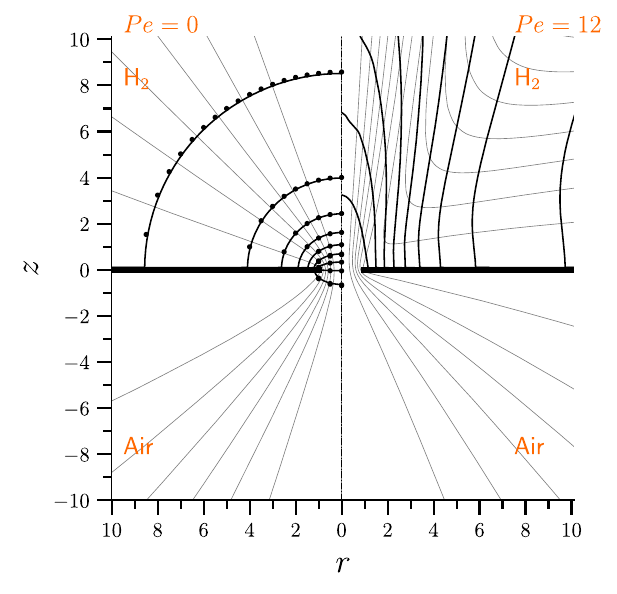}
    \caption{Streamlines $\psi=$~constant (gray curves, streamfunction increment $\Delta \psi = 0.025$) and isocontours of mass fraction $Y=$~constant (black curves, ${\Delta Y = 0.1}$) calculated for Air-H$_2$ with $Pe = 0$ (left) and $Pe=12$ (right). The dots on the left half represent the theoretically predicted mass-fraction isocontours given in Eq.~\eqref{paraboloidal surfaces}.}
    \label{fig:Air-H2_iso}
\end{figure}

The finite element method was used to integrate the problem after writing Eqs.~\eqref{cont}–\eqref{Yeq} in weak form. Piecewise differentiable linear, quadratic, and cubic shape functions were used for the pressure, velocity, and mass fraction, respectively, all vanishing at the edge of the computational domain, $\bm{x} = (r, z) \in \partial V$. The problem was discretized on an unstructured triangular mesh using Taylor–Hood elements. Element size was refined near the orifice and the wall, reaching a minimum value of $\delta = 0.01$, and gradually increased with distance to reach a maximum size of $\delta = 19.63$ at the outer boundary. Boundary conditions at $r^2 + z^2 \to \pm\infty$ were imposed on a truncated numerical domain defined by $r^2 + z^2 \to R^2$, with $R \simeq 250$ (case dependent). The computations were performed using COMSOL Multiphysics \cite{multiphysics2019comsol}, employing a fully coupled iterative Newton method \cite{deuflhard1974modified}. Iterations were continued until the weighted Euclidean norm of the residual vector fell below $10^{-6}$.

\begin{figure}[!ht]
    \centering
    \includegraphics[width=\linewidth]{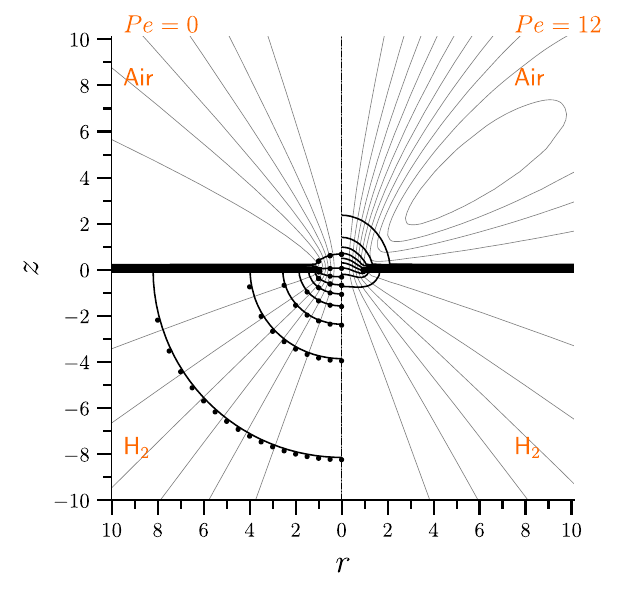}
    \caption{Streamlines $\psi=$~constant (gray curves, streamfunction increment $\Delta \psi = 0.025$) and isocontours of mass fraction $Y=$~constant (black curves, ${\Delta Y = 0.1}$) calculated for H$_2$-Air with $Pe = 0$ (left) and $Pe=12$ (right). The dots on the left half represent the theoretically predicted mass-fraction isocontours given in Eq.~\eqref{paraboloidal surfaces}.}
    \label{fig:H2-Air_iso}
\end{figure}

A total of $n_0 = 2000$ elements were used in the computations presented below. To verify grid independence, we monitored the Sherwood number while varying $R$, the number of elements $n$, and the minimum element size $\delta$. Variations in the Sherwood number remained below $\sim 10^{-3}$ when the size of the domain $R$ was increased keeping the minimum size of the elements and when $n$ was doubled keeping the domain size constant. The Air-H$_2$ gas pair with $Pe=12$ was chosen for the grid-independency study and to test the influence of the domain size on the solution.

\subsection{The velocity and concentration fields}

Figures~\ref{fig:sketch}--\ref{fig:H2-Air_iso} show streamlines (thin gray solid curves) and isocontours of mass fraction (thick black solid curves) determined numerically for different values of $\Pe$. The streamlines are visualized by plotting the isocontours of the stream function $\psi$, evaluated from
\beq
\rho r v_r=-\frac{\p \psi}{\p z} \quad {\rm and} \quad \rho r v_z=\frac{\p \psi}{\p r},
\eeq
with $\psi=0$ along the axis.

\begin{figure}[!ht]
    \centering          
        \centering
        \includegraphics[width=\linewidth]{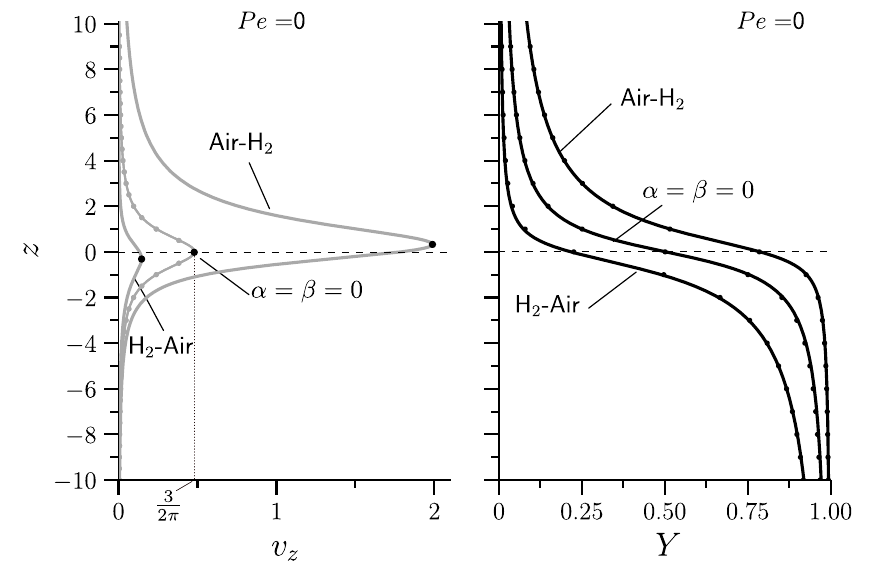}               
    \caption{Numerically computed distributions of axial velocity $v_z$ (left) and mass fraction $Y$ (right) along the symmetry axis $r=0$ for $Pe=0$ and different pairs of gases (indicated in the figure). The dots in the right-hand-side plot corresponds to the analytical solution given in Eq. \ref{Yz_sampson}, while those in the left-hand-side plot represent the constant-density solution obtained by Sampson \cite{sampson1891}, given in Eq.~\eqref{vz_sampson}. The black circles in the left figure indicate the location at which the maximum velocity is achieved.}
    \label{fig:comparison_Pe0}
\end{figure}

\begin{figure}[!ht]
    \centering          
        \centering
        \includegraphics[width=\linewidth]{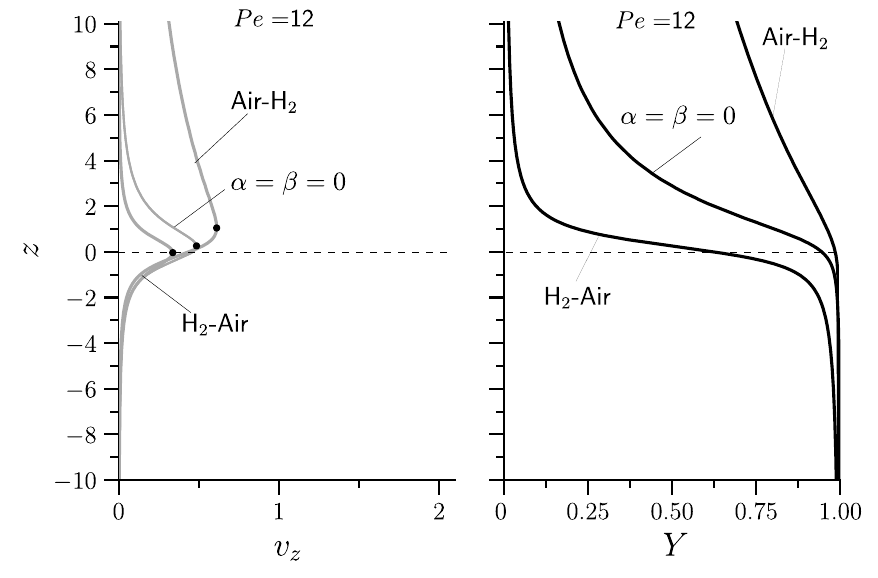}                   
   \caption{Numerically computed distributions of axial velocity $v_z$ (left) and mass fraction $Y$ (right) along the symmetry axis $r=0$ for $Pe=12$ and different pairs of gases (indicated in the figure). The black circles in the left figure indicate the location at which the maximum velocity is achieved.}
    \label{fig:comparison_Pe12}
\end{figure}
Sample results corresponding to creeping flow ($\Pe=0$) are shown in Fig.~\ref{fig:sketch}. For the case $\alpha=\beta=0$ considered, the solution reduces to one with constant density and viscosity, for which analytical expressions are available for both the velocity field \cite{sampson1891} and the mass fraction \cite{atwal2022mass}. The streamlines computed numerically are compared with the exact constant-density solution $\psi=(1-\sin^3 \eta)/(2\pi)$ obtained by Sampson \cite{sampson1891}, in which the stream surfaces reduce to the hyperboloids
\beq \label{psi_sampson}
\frac{r^2}{1-(1-2\pi \psi)^{2/3}}-\frac{z^2}{(1-2\pi \psi)^{2/3}}=1
\eeq
with $0 \le \psi \le 1/(2\pi)$. The figure also compares the numerically computed isocontours of mass fraction with the constant-density analytical solution of Atwal et al. \cite{atwal2022mass}, which yields the paraboloidal isosurfaces
\beq \label{Y_sampson}
\sin^2[\pi(1-Y)] r^2+ \tan^2[\pi(1-Y)] z^2=1,
\eeq
as follows from Eq.~\eqref{paraboloidal surfaces} when $\alpha \ll 1$. The perfect agreement demonstrated by the comparison lends confidence to the accuracy of the numerical method.

The flow structure for the gaseous configurations is examined in Figs.~\ref{fig:Air-H2_iso} and~\ref{fig:H2-Air_iso}, which present results for air–H$_2$ and H$_2$–air at $\Pe = 0$ and $\Pe = 12$. It is instructive to focus first on the left halves of the figures, corresponding to the creeping-flow case $\Pe = 0$. As can be seen, the isocontours of mass fraction give perfect agreement with the analytic predictions of Eq.~\eqref{paraboloidal surfaces}, which are represented with the small circles in the figure. The plots reveal that, despite the absence of inertia, the variation of both density and viscosity with concentration breaks the flow symmetry about the plane of the orifice plate. This behavior stands in stark contrast to the constant-property case ($\alpha = \beta = 0$) illustrated in Fig.~\ref{fig:sketch}, where perfect symmetry is preserved.

Another notable feature is the asymmetry in the concentration field. Because the diffusive flux is linearly proportional to the density, as reflected in the advection-free transport equation $\bnabla \cdot (\rho \bnabla Y)=0$, the concentration gradient is significantly steeper on the hydrogen side of the orifice, where the density is smaller, whereas relative variations in composition remain modest on the air side.

These effects become more evident in Fig.~\ref{fig:comparison_Pe0}, which shows the variation of vertical velocity $v_z$ and mass fraction $Y$ along the symmetry axis $r = 0$ for both air–\Hy and \Hy–air with $\Pe=0$, alongside the corresponding constant-property profiles. The plots clearly display the asymmetry of the solution and the markedly different species gradients on either side of the orifice. As expected for purely diffusive species transport, the mass-fraction profiles satisfy the relation $Y_{\rm H_2\text{-}Air}(z) + Y_{\rm Air\text{-}H_2}(-z) = 1$; that is, the composition field remains unchanged when the roles of the high- and low-pressure gases are reversed, provided the flow occurs at low Péclet number. In both cases, the mass fraction of hydrogen at the orifice is $Y_{\rm H_2} \simeq 0.21$, as can be obtained by using Eq.~\eqref{eq:Y_o} to evaluate $Y_o$ (for H$_2$-Air with $\alpha=-0.93$) or $1-Y_o$ (for Air-H$_2$ with $\alpha=13$).

To further validate the numerical results, the velocity profile for the constant-property case ($\alpha = \beta = 0$) with $\Pe=0$ is compared in the left plot of Fig.~\ref{fig:comparison_Pe0} with Sampson’s constant-density solution \cite{sampson1891}
\beq \label{vz_sampson}
v_z=\frac{3}{2 \pi (z^2+1)},
\eeq
and the mass-fraction distributions are compared in the right plot with the analytical predictions
\beq \label{Yz_sampson}
Y=1-\frac{1}{\alpha} \left\{\exp\left[\frac{\ln(1+\alpha) }{\pi}\tan^{-1}\left(z^{-1}
\right)\right]-1 \right\}.
\eeq
stemming from Eq.~\eqref{paraboloidal surfaces} when $r=0$, with $\alpha$ given in Table~\ref{table}.


\begin{figure}[!ht]
    \centering
    \small    
    \includegraphics[width=\linewidth]{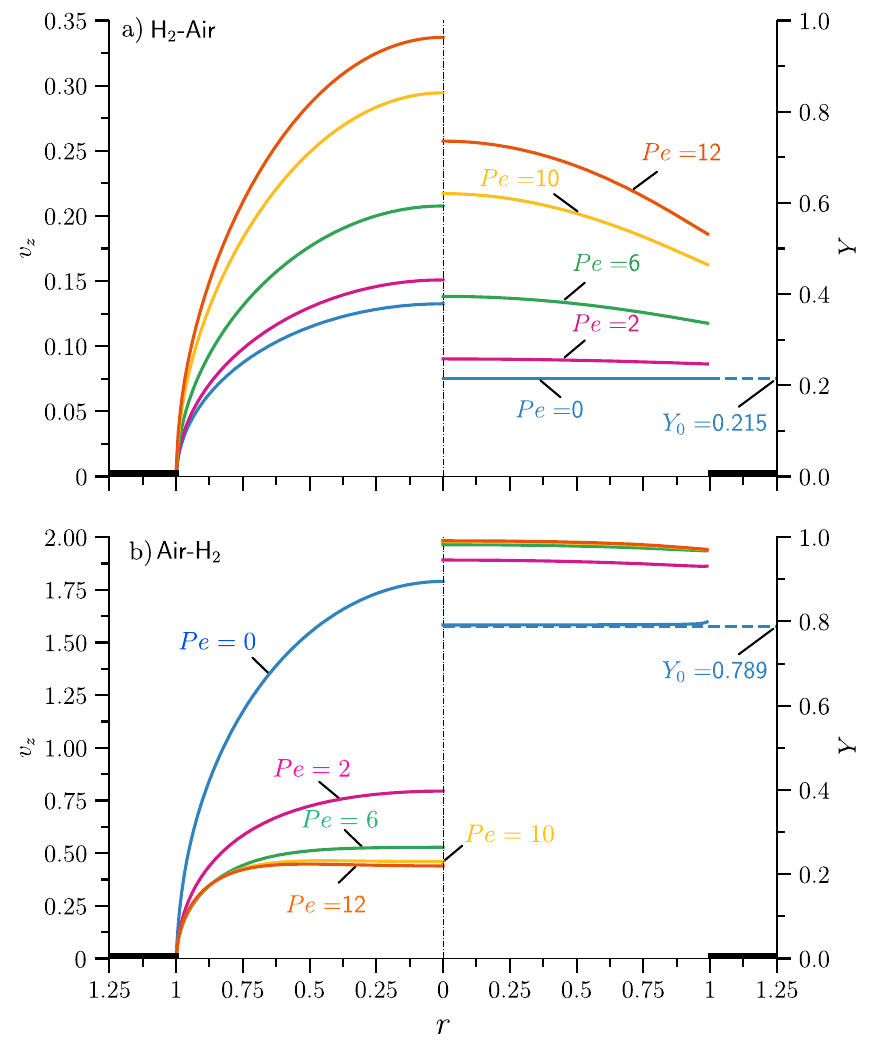}    
    \caption{Numerically computed distributions of axial velocity $v_z$ (left) and mass fraction $Y$ (right) across the orifice for hydrogen discharging into air (a) and air discharging into hydrogen (b) at different P\'eclet numbers (indicated in the figure). Notice that the color code in the figure corresponds to the values of $\Pe$ indicated by the labels. The dashed lines indicate the value of the mass fraction at the orifice given in~Eq.~\eqref{eq:Y_o} for $\Pe=0$.} 
    \label{fig:radial_profiles}
\end{figure}


The results for $\Pe = 12$, shown on the right-hand sides of Figs.~\ref{fig:Air-H2_iso} and~\ref{fig:H2-Air_iso}, demonstrate that as the Péclet number increases to moderately large values, advection becomes dominant and the concentration near the orifice approaches $Y = 1$, as expected. \review{The streamlines reveal that the flow begins to approach the large--Reynolds-number structure predicted for constant-density jets by Schneider \cite{schneider1981flow}, featuring a jet of nearly constant momentum flux surrounded by a low-velocity viscous flow driven by jet entrainment. In this context, the recirculatory pattern visible on the right-hand side of Fig.~\ref{fig:H2-Air_iso} bears a striking resemblance to that obtained in computations of submerged water jets at $\Reyn=5.5$ \cite{schneider1987recirculatory}.}

Advection effects are more pronounced when air discharges into hydrogen, \review{the case shown in Fig.~\ref{fig:Air-H2_iso}, than when hydrogen discharges into air, the case shown in Fig.~\ref{fig:H2-Air_iso}, with the streamlines in the latter case diverging} more readily as the gas crosses the orifice, reflecting the rapid axial velocity decay characteristic of slender light jets \cite{sanchez2011variable}. These differences are clearly reflected in the axial profiles presented in Fig.~\ref{fig:comparison_Pe12}. For the H$_2$–air case, both velocity and mass fraction decay rapidly for $z > 0$, whereas in the air–H$_2$ case, the decay is much more gradual. This latter behavior is consistent with theoretical predictions previously established for heavy jets at moderately large Reynolds numbers~\cite{rosales-vera-2016}.

To complete the characterization of the flow, radial profiles of $v_z$ and $Y$ at the orifice plane ($z = 0$) are shown in Fig.~\ref{fig:radial_profiles} for the Air–H$_2$ and H$_2$–Air configurations at different values of $\Pe$. In the former case, the transverse diffusion of hydrogen into the dense air jet emerging from the orifice is minimal, keeping the mass fraction $Y$ nearly constant for $r < 1$. Its uniform value rapidly approaches $Y = 1$ as $\Pe$ increases. The evolution is markedly different for hydrogen discharging into air. As shown in the figure, the mass fraction remains uniform and equal to $Y = 0.209$ for $\Pe = 0$, in agreement with Eq.~\eqref{eq:Y_o}. As $\Pe$ increases, advection enhances the hydrogen content, with diffusion from the surrounding air playing a lesser role near the axis. As a result, the hydrogen concentration becomes higher along the centerline, producing a parabolic-like mass-fraction profile. Even for the largest value considered ($\Pe = 12$), the mixture across the orifice contains more than 20\% air.

The velocity distributions, shown on the left-hand side of Fig.~\ref{fig:radial_profiles}, evolve from a parabolic-like shape at $\Pe = 0$ to a progressively flatter profile near the axis as $\Pe$ increases. This flattening is more pronounced for air discharging into hydrogen. For even larger $\Pe$ values, not considered here, the solution is expected to approach the inviscid limit, featuring a sharp interface bounding a jet of gas 1. The jet-contraction ratio, which takes the familiar value 0.61~\cite{Trefftz1917Uber} for water jets discharging into air, is expected to depend on the gas-pair density ratio. In this $\Pe \gg 1$ regime, mixing occurs downstream of the orifice as gas 1 entrains the surrounding gas 2. For hydrogen discharging into air, a detailed high-Péclet-number numerical description of the downstream jet development region is available in~\cite{sanchez2010hydrogen}.

\subsection{The Sherwood number}

\begin{figure}[!ht]
    \centering
    \includegraphics[width=\linewidth]{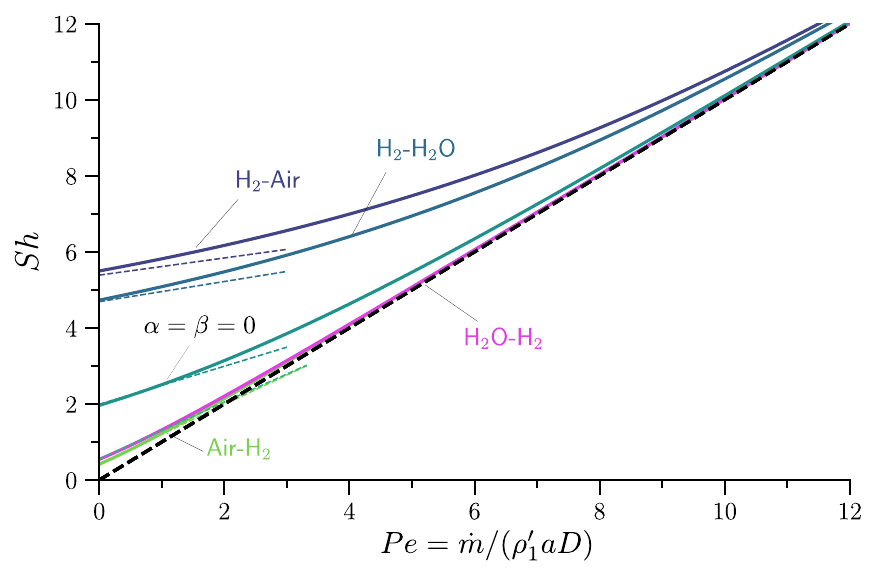}
    \caption{The variation of the Sherwood number $Sh$ with $Pe$ for several gas pairs. The thin dashed lines represent the asymptotic prediction given by Eq.~\eqref{eq:Sh} for $Pe \ll 1$. The thick dashed line represents $Sh=Pe$. } 
    \label{fig:Sh_vs_Pe}
\end{figure}

The Sherwood number defined in Eq.~\eqref{Sh_def} quantifies gas transport across the orifice, with the mass fluxes of the two gases related by
\beq \label{m1_m2}
\frac{\dot{m}_1}{\Sh} = \frac{\dot{m}_2}{\Pe - \Sh} = \rho'_1 a D.
\eeq
Its numerically computed dependence on the Péclet number is shown in Fig.~\ref{fig:Sh_vs_Pe} for the gas pairs listed in Table~\ref{table}, as well as for the constant-property case $\alpha = \beta = 0$ previously analyzed in \cite{atwal2022mass}. The curves corresponds to the numerical solution of the problem for $0 \le Pe \le 12$ with increments on the Peclet number $\Delta Pe=0.1$.
The dotted lines represent the linear asymptotic expression given in Eq.~\eqref{eq:Sh}. As can be seen, although this approximation is strictly valid only in the limit $\Pe \ll 1$, it remains reasonably accurate for Péclet numbers of order unity, with deviations from the numerical results remaining below approximately 10\% for $\Pe \le 3$.

For reference, Fig.~\ref{fig:Sh_vs_Pe} also shows the straight line $\Sh = \Pe$. As indicated by Eq.~\eqref{m1_m2}, the difference $\Pe - \Sh < 0$ between this reference value and the computed Sherwood number determines the mass flux of gas 2, with the negative sign indicating transport in the negative $z$ direction—that is, toward the region of higher pressure. As expected, the Sherwood number approaches the Péclet number in the limit $\Pe \gg 1$, corresponding to $\dot{m}_1 \rightarrow \dot{m}$ and $\dot{m}_2 \rightarrow 0$. The curves for Air–H$_2$ and H$_2$O–H$_2$, which are nearly indistinguishable, lie close to the line $\Sh = \Pe$ across all values of $\Pe$. This behavior reflects the limited ability of the light gas to diffuse into the heavy gas, as also evident in the flow structures shown in Figs.~\ref{fig:Air-H2_iso} and~\ref{fig:H2-Air_iso}.

\begin{figure}[!ht]
    \centering
    \includegraphics[width=\linewidth]{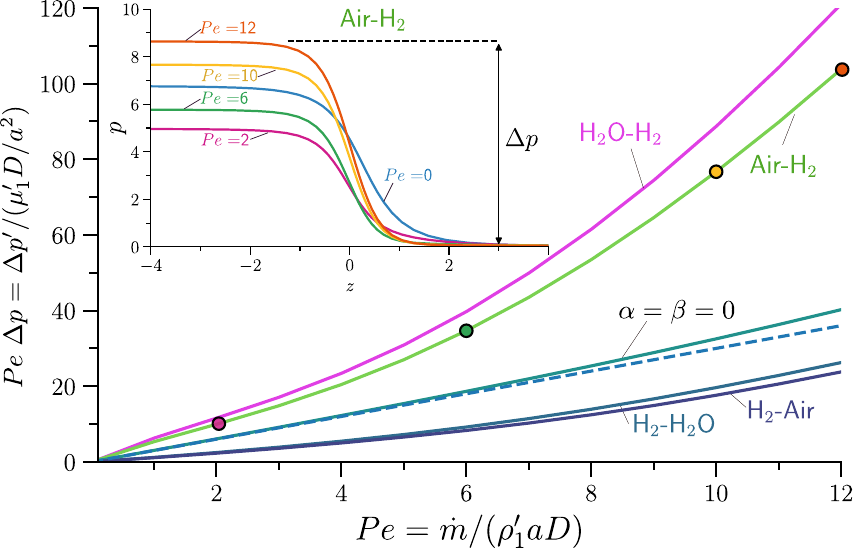}
    \caption{The numerically computed variation of the dimensionless intramural pressure difference $\Delta p' / (\mu_1' D / a^2)$ with $\Pe$ for different gas pairs. The dashed line represents the curve $\Delta p'/(\mu D/a^2)=3 Pe$ corresponding to Sampson's constant-density creeping-flow solution. The inset includes the  pressure profiles along the symmetry axis $r=0$ for Air-\Hy.}
    \label{fig:Delta_p}
\end{figure}

\subsection{Pressure drop across the orifice}

The integration of Eqs.~\eqref{cont}--\eqref{Yeq} for a given value of $\Pe$ yields the dimensionless pressure field $p(r,z)$ as part of the solution. Along the axis, the pressure decreases from $p = \Delta p$ as $z \rightarrow -\infty$ to $p = 0$ as $z \rightarrow \infty$, where $\Delta p$ denotes the dimensionless pressure difference required to sustain the flow rate $\dot{m}$. This quantity is related to the dimensional pressure difference $\Delta p'$ by $\Delta p = \Delta p'/[\mu_1' \dot{m} / (\rho_1' a^3)]$, as follows from Eq.~\eqref{p_def}. 

For the special case of constant-property creeping flow \cite{sampson1891}, the axial pressure distribution is given analytically by $p=(3/\pi)[\pi/2-\tan^{-1} z-z/(1+z^2)]$ with the associated pressure drop reducing to $\Delta p = 3$. In all other cases, both the pressure profile and the corresponding value of $\Delta p$ must be computed numerically. Sample results are shown in the inset of Fig.~\ref{fig:Delta_p} for the case of air discharging into hydrogen.

The dependence of the pressure drop on the Péclet number is shown in Fig.~\ref{fig:Delta_p}. For plotting purposes, we define the dimensionless quantity $\Pe \Delta p = \Delta p' / (\mu_1' D / a^2)$, which is independent of the mass flow rate $\dot{m}$. With this representation, the curves in Fig.~\ref{fig:Delta_p} directly provide the pressure difference $\Delta p'$ needed to generate a given mass flow rate $\dot{m}$ for a specific gas pair and orifice size. Results are provided for the gas pairs listed in Table~\ref{table}. To demonstrate the differences with the case of liquid mixing, the figure also includes the numerical solution corresponding to the constant-property case $\alpha = \beta = 0$, along with the asymptotic prediction $\Delta p'/(\mu D/a^2) = 3 \Pe$ corresponding to Sampson’s creeping-flow solution \cite{sampson1891}, shown as a dashed line.



\section{Concluding remarks}

This study presents a detailed quantitative investigation of mass transport in binary gas mixtures through an orifice, focusing on the distinguished intermediate regime in which advection and diffusion are of comparable magnitude; characterized by order-unity values of the Schmidt number $\Schm$ and Péclet number $\Pe$. The analysis accounts for significant variations in density and viscosity across the mixture. Owing to the coupling between the velocity and concentration fields, numerical integration is generally required to determine the Sherwood number $\Sh$, which characterizes the mass transport. An exception occurs in the purely diffusive limit $\Pe \ll 1$, where a closed-form solution has been derived by formulating the transport problem in terms of oblate spheroidal coordinates.

Numerically determined flow structures for hydrogen discharging into air and air discharging into hydrogen are examined to highlight the effects of fluid-property variations. Diffusion is found to be more pronounced on the side of the orifice where the gas density is lower. As the Péclet number $\Pe$ increases, a jet begins to form, becoming significantly more robust when the heavier gas discharges into the lighter one—an expected outcome due to the enhanced inertia.

The main quantitative results of the study are summarized in Figs.~\ref{fig:Sh_vs_Pe} and~\ref{fig:Delta_p}. Figure~\ref{fig:Sh_vs_Pe}, which shows the variation of the Sherwood number with $\Pe$ for hydrogen–air and hydrogen–water vapor mixtures, can be used in conjunction with Eq.~\eqref{m1_m2} to evaluate the individual mass fluxes of the two gases, while figure~\ref{fig:Delta_p} relates the transmural pressure difference $\Delta p'$ to the overall mass flux $\dot{m}$.

\review{As an illustrative application, consider the flow of hydrogen into air through a small orifice of radius $a = 400~\mu\mathrm{m}$. We seek to determine the overpressure $\Delta p'$ required to sustain a hydrogen mass flux ${\dot{m}_{\scriptscriptstyle{\rm H_2}} = 1}$ {mg/min}. This mass flux corresponds to a Sherwood number ${\Sh = 7.873}$, obtained by evaluating the first equation in~\eqref{Sh_def} using $D = 0.630 \times 10^{-4}~\mathrm{m}^2/\mathrm{s}$ and $\rho'_{\scriptscriptstyle{\rm H_2}} = 0.084~\mathrm{kg/m}^3$, appropriate for hydrogen at $p_o = 1~\mathrm{atm}$ and $T_o = 293~\mathrm{K}$. The associated P\'eclet number, $\Pe \simeq 5.940$, is inferred from the curve labeled H$_2$--air in Fig.~\ref{fig:Sh_vs_Pe}. The mass flux of air follows from the first equation in~\eqref{m1_m2}, yielding $\dot{m}_{\scriptscriptstyle{\rm Air}}=\dot{m}_{\scriptscriptstyle{\rm H_2}} (\Pe-\Sh)/\Sh \simeq -0.245$ mg/min, where the negative sign indicates that air flux is into the hydrogen vessel.  From the curve labeled H$_2$--air in Fig.~\ref{fig:Delta_p}, the value $\Delta p'/(\mu'_{\scriptscriptstyle{\rm H_2}} D/a^2) =9.640$ is obtained for $\Pe \simeq 5.94$. It follows that $\Delta p'=33.403$ mPa, where the hydrogen viscosity $\mu'_{\scriptscriptstyle{\rm H_2}} = 8.81 \times 10^{-6}~\mathrm{kg/(m\,s)}$ has been used.} 

\review{In configurations where the pressure difference $\Delta p'$ is prescribed, the computational procedure proceeds in reverse. One first uses Fig.~\ref{fig:Delta_p} to determine the P\'eclet number $\Pe$ and the total mass flow rate $\dot{m} = \Pe\, \rho_{\scriptscriptstyle{\rm H_2}}' a D$ as a function of $\Delta p'/(\mu'_{\scriptscriptstyle{\rm H_2}} D / a^2)$. The resulting value of $\Pe$ is then used in Fig.~\ref{fig:Sh_vs_Pe} to evaluate the Sherwood number $\Sh$, from which the mass flux of hydrogen follows as $\dot{m}_{\scriptscriptstyle{\rm H_2}} = \Sh\, \rho_{\scriptscriptstyle{\rm H_2}}' a D$. The flux of the air is finally obtained by difference, $\dot{m}_{\scriptscriptstyle{\rm Air}} = \dot{m} - \dot{m}_{\scriptscriptstyle{\rm H_2}}$.}


The methodology developed in this study provides a comprehensive framework for predicting and controlling the transport characteristics of binary gas mixtures through small orifices, with direct relevance to engineering systems involving gas flow at moderately low Reynolds numbers. Although illustrative results are presented only for H$_2$–air and H$_2$–H$_2$O mixtures, the approach can be readily extended to other binary gas combinations, offering quantitative insights applicable to a range of technologies, including semiconductor manufacturing, gas metering, and safety systems.

\section*{Acknowledgments}

We thank Prof. Stefan Llewellyn Smith for fruitful discussions on low-Reynolds-number flows and for bringing to our attention the seminal work of Weber (1873) and the reference book by Sneddon (1966). The senior author (ALS) gratefully acknowledges Jeffrey Spiegelman, an entrepreneur with extensive experience in water vapor purification and delivery for oxidation processes in semiconductor and microelectronics fabrication, for enlightening discussions on orifice design for gaseous flow control that helped motivate this study. MSS gratefully acknowledges the support of project PID2022-139082NB-C51  and Salvador de Madariaga-Fullbright  program (PRX22/00031) funded by MCIN/AEI, Spain.  

\bibliography{orifice_bibliography}

\end{document}